\documentstyle[sprocl]{article}

\input{psfig}

\def\be{\begin{equation}}
\def\ee{\end{equation}}
\def\bea{\begin{eqnarray}}
\def\eea{\end{eqnarray}}

\begin{document}

\title{THREE NUCLEONS AT VERY LOW ENERGY}

\author{PAULO  F. BEDAQUE}

\address{Institute for Nuclear Theory\\
         University of Washington,\\
         Seattle, WA 98195}


\maketitle\abstracts{We discuss the effective field theory approach to
nuclear phenomena with typical momentum below the pion mass.
We particularly focus on the three body problem. In the $J=3/2$ channel,
effective field theory is extremelly predictive and we are able to describe
the nucleon-deuteron scattering amplitude at the percent level using only 
two parameters 
obtained from nucleon-nucleon scattering. We briefly comment on the $J=1/2$ 
channel and the issues related to the non-perturbative renormalization of 
the three body force.}

\section{Introduction}

Most of the early work on the use of effective field theory in nuclear 
physics concentrated on the momentum scale $p \sim m_\pi$. There is a good 
reason for that: most of nuclear phenomenology is at this scale (the 
Fermi momentum in nuclear matter, for instance, is at the order of $2 m_\pi$).
This talk, however, will concentrate on the scale $p \sim 1/a \ll m_\pi$ ($a$
is the nucleon-nucleon scattering lenght), 
that will 
be refered to as the `` very low '' momentum scale. This scale region 
is interesting for three reasons. 
The first reason is that 
this scale,
even though not typical in nuclei, {\it is} relevant for a  corner 
of nuclear physics. This comes about because
the deuteron is an anomalously shallow bound state, with a binding energy 
of the order of $1/M a^2$, what sets the momentum scale of deuteron 
physics to be $p\sim 1/a\ll m_\pi$.
A second reason is that, since pions can be integrated out
and don't appear as explicit degrees of freedom, a number of issues related to
regularization and renormalization, many of them extensively discussed in 
this workshop, can be addresed in a simpler situation.  
The third motivation to consider the `` very low '' 
momentum scale is related to recent work on the  ``  low '' momentum scale 
$p \sim m_\pi$ . At this higher momentum scale pions have to be included 
explicitly and in previous power counting schemes the whole ladder of pion 
exchanges contributed at leading order to nucleon-nucleon scattering. This 
is a source of tremendous complication, both conceptual and practical. 
Recently  a new power counting scheme was 
suggested \cite{ksw},\cite{martin} in which
pion exchanges are perturbative and, in particular, the leading order 
contains no pion exchanges. Thus, all results obtained discussed here 
using the 
effective theory appropriate to the $p \ll m_\pi$  scale will 
automatically equal  the leading order result obtained from this new
 power counting of  valid at the $p \sim m_\pi$ scale. 

In the absence of explicit pions, the effective theory method applied to 
nucleon-nucleon scattering simply reproduces the effective range expansion.
This is phenomenologically correct, but hardly interesting. The first 
nontrivial application is the three body problem, that is the main subject 
of this talk.

\section{The very low energy effective theory}

For momenta much smaller than the pion mass, the only  degrees of 
freedom that need to be included explicitly
are the nucleons. The effect of pion exchanges, $\Delta$'s, heavier meson, 
etc., are all implicitly included in the coefficient of local operators in 
the lagrangean. We will concentrate later on the three body problem
in the $J=3/2$ channel. In this channel all nucleon spins are parallel thus 
all s-wave two nucleon interactions are in the spin triplet channel. 
Restricting
ourselves to spin triplet interactions, the most general lagrangean is :

\begin{eqnarray}
{\cal L}&=&  N^\dagger(i\partial_{0}+\frac{\vec{\nabla}^{2}}{2M}+\ldots)N 
         + C_0 (N^\dagger \tau_2\vec{\sigma}\sigma_2 N)^2\nonumber \\ 
        &&+C_2 \left[ (N^\dagger \tau_2\vec{\sigma}\sigma_2\nabla N)
               (N^\dagger \tau_2\vec{\sigma}\sigma_2\nabla N) 
            -3(N^\dagger \tau_2\vec{\sigma}\sigma_2 N)
             (N^\dagger \tau_2\vec{\sigma}\sigma_2 \nabla^2 N)+h.c.\right]
               \nonumber\\
            && +\ldots, \label{lag}
\end{eqnarray}
where $M$ is the nucleon mass, $C_n$ are constants related to the two-body 
force terms containing $n$ derivatives, and the dots stand for higher-order
terms including relativistic corrections, 
higher-derivative terms, 
three-body forces, etc. Terms describing the spin singlet interactions
have a similar form but will not be needed here.
The constants $C_n$ are determined by nucleon-nucleon scattering data.
As it was explained in more detail in van Kolck's \cite{bira} 
talk it turns out that, 
using dimensional regularization and minimal subtraction,
$C_0\sim a/M$, $C_2\sim r_0 (r_0 a)/M$, $C_4\sim r_0 (r_0 a)^2/M+\ldots$
and so on ($r_0$ is the effective range and ellipses stand for terms 
suppressed by powers of $r_0/a$ ).
The leading pieces in each one of these terms form a geometric series that
can be conveniently summed to all orders by the introduction of a field 
of baryon-number two \cite{transvestite}

\begin{eqnarray}
\cal L & = & N^\dagger(i\partial_{0}+\frac{\vec{\nabla}^{2}}{2M}+\ldots)N 
         + \vec{d}^\dagger\cdot(-i\partial_{0}-\frac{\vec{\nabla}^{2}}{4M}
                             +\Delta+\ldots)\vec{d} \nonumber \\ 
 &  & -\frac{g}{2} (\vec{d}^\dagger\cdot N\vec{\sigma}\sigma_2\tau_2 N 
                       +\mbox{h.c.})
 +\ldots                       \label{lagt}
\end{eqnarray}
\noindent
If the dibaryon field $\vec{d}$ is integrated out,
the Lagrangian (\ref{lag}) is recovered as long as $\Delta$ and $g$ are 
appropriate functions of $C_0$ and $C_2$. 
This resummation is by no means necessary, since for momenta of the order 
$p\sim 1/a$ the resummed terms are subleading, but it is just a convenient
way of computing higher-order corrections.

\begin{figure}[htb]
\begin{center}
\psfig{figure=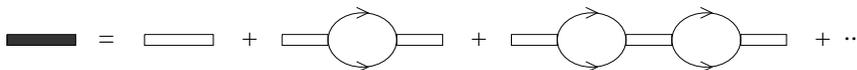,width=4.5in}
\end{center}
\caption{Dressed dibaryon propagator. }
\label{fig1}
\end{figure}

The numerical values of $g$ and $\Delta$ can be determined if we 
consider the dressed dibaryon propagator (Fig. \ref{fig1}).

\vskip 0.3in
\noindent
The linearly divergent loop integral is set to zero in dimensional 
regularization and the result is
\begin{equation}
i S(p) =  \frac {1}{p^0- \frac{\vec{p}^{\,2}}{4M} - \Delta
             + \frac{M g^{2}}{2\pi} 
               \sqrt{-M p^0+\frac{\vec{p}^{\,2}}{4}-i\epsilon} +i\epsilon} .
                                   \label{Dprop}
\end{equation}
\noindent
This propagator is, up to a constant, the scattering matrix of two nucleons
in the $^3S_1$ channel,
\begin{equation}
T(k) = {4 \pi \over M} 
                    {1\over -\frac{2 \pi \Delta}
                                  {M g^{2}} 
                    +\frac{2 \pi }
                          {M^2 g^{2}}k^2
                    -i k},      \label{NNamp}
\end{equation}
\noindent
where $k^2/M$ is the energy in the center-of-mass frame. 
This result is just the 
familiar effective range expansion, from what we can infer the proper 
values for the 
constants $g$ and $\Delta$. Using  $a= 5.42$ fm and 
$r_0=1.75$ fm \cite{nijm},
we find
\begin{eqnarray}
g^2    &= \frac{4 \pi}{M^2 r_0}  &= 1.6 \cdot 10^{-3} \ {\rm MeV^{-1}},\label{g}\\
\Delta &= \frac{2}{M a r_0}    &= 8.7 \ {\rm MeV}\label{Delta}. 
\end{eqnarray}
{}From Eqs. (\ref{Dprop}), (\ref{g}), and (\ref{Delta}) we see why it
is necessary to resum the bubble graphs in Fig. \ref{fig1} to all orders
for $p \sim 1/a$: the term in the square root coming from the unitarity
cut is of the same order as $\Delta$. On the other hand, as mentioned
before, the kinetic term of the dibaryon is smaller than the other terms
in (\ref{Dprop}) and is resummed for convenience only.
Notice that the propagator (\ref{Dprop}) has two poles, one at
$p^0= \vec{p}^{\,2}/4M-B$ (the deuteron pole), another at
 $p^0= \vec{p}^{\,2}/4M -B_{deep}$ (unphysical deep pole), 
and a cut along the positive real axis
starting at $p^0= \vec{p}^{\,2}/4M$.

\section{Power counting for  neutron-deuteron scattering}

 Let us now turn  to neutron-deuteron scattering. 
We want to identify which graphs give the leading contributions in an expansion
on powers of the typical momentum $p\sim 1/a$ over the scale of the physics
not explicitly included in the effective theory $m_\pi\sim 1/r_0$.

\begin{figure}[htb]
\begin{center}
\psfig{figure=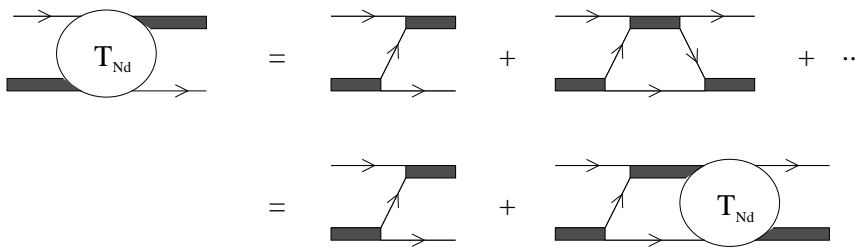,width=4.5in}
\end{center}
\caption{Dressed dibaryon propagator. }
\label{fig2}
\end{figure}

The simplest 
diagram contributing to neutron-deuteron scattering is the first diagram on 
the right hand side of 
Fig. \ref{fig2}.
\vskip .3in

{}For momenta of the order of $p \sim 1/a$ it gives a contribution 
of the order of $ g^2/(p^2/M) \sim a^2/M r_0$. The one-loop graph
mixes different orders of the expansion, since it
involves the dibaryon propagator 
$g^2/(\Delta+p^2/M)\sim (a/M)(1+{\cal O} (r_0/a) +\ldots)$.
It also involves a loop. The contribution coming from loops can be 
estimated by rescaling all internal energies by $p^2/M$, ,where $p$ is some
typical external momentum. The result is $\sim M^n p^m$, where $n=$number 
of nucleons propagators minus the number of loops and $m$ can be determined 
by dimensional analysis. Using this rule we can easily estimate the one loop 
graph in Fig.~\ref{fig2} to be $g^2 g^2/(\Delta+p^2/M)\sim 
(a^2/M r_0) (1 + {\cal O} (r_0/a) + \ldots)$. 
Similarly we can see that the remaining graphs in 
Fig. \ref{fig2} 
give contributions of the same order, which means that an infinite number
of diagrams contribute to the leading orders. 

Other contributions are suppressed by at least three powers of 
$r_0/a$ or $p r_0$. 
For instance, the effect of the subleading (not resummed)
piece of $C_4$  is to generate
the shape parameter ($\sim k^4$) term 
in the effective range expansion of the nucleon-nucleon interaction. 
Its typical size is
$\sim k^4 r_0^3$ compared to the leading piece $\sim 1/a$ and is thus also 
suppressed by $(r_0/a)^3$. 
Likewise, p-wave interactions, unaffected by the existence
of a shallow s-wave bound state, arise from a term in 
the lagrangean with two derivatives and a coefficient of the order 
$\sim 1/M m_\pi^3$.
We conclude then that a  diagram made out of the substitution of 
one of the dibaryon 
propagators in a diagram in 
Fig. \ref{fig2} by a p-wave interaction vertex would be suppressed by
$(r_0/a)^3$ in comparison to the leading order. 
Three-body force terms have to contain at least 
two derivatives since in the $J=3/2$ channel all the spins are up and 
Fermi statistics forbids the placement of all three nucleon in a s-wave. The
natural size of the coefficient of the six nucleon, two derivative
term is $1/M m_\pi^6$. A diagram including this three body force includes
two nucleon loops, each one giving a $M p$ factor, two dibaryon-two nucleons
vertices and a three body force vertex, for a total contribution of the 
order of $p^4/M m_\pi^6 r_0 \sim a^2/M r_0 (r_0/a)^6$.
Thus contributions coming from 
the three-body force are suppressed in relation to the leading order 
graphs
by 
$(r_0/a)^6$. 
\section{Summing the leading graphs}
{}From the results of the previous section we see that a 
 calculation accurate up to corrections of order $(r_0/a)^3$ is possible
by summing the diagrams of Fig. \ref{fig2}. Fortunately, the interaction 
mediated by the s-channel dibaryon generates a very simple, local and separable
potential between nucleons. It is well known that the three-body problem
with separable two-body interactions reduces to an equivalent two-body problem.
In our case the equation to be solved can be read off Fig. \ref{fig2},
and an integration over the energy inside the loop gives \cite{genius}

\begin{eqnarray} 
 \tilde t(\vec{p}, \vec{k})
 =& -&\frac{1}
        {(\vec{p}-\vec{k}/2)^2+M B}\nonumber \\ 
  & -& \int \frac{d^3 l}
                 {(2\pi)^3}
       \frac{\tilde t(\vec{l}, \vec{k})} 
            {-\frac{3(\vec{l}^{\, 2}-\vec{k}^2) }
                   {8 M^2 g^2}
              +\frac{1}
                    {4 \pi}({\sqrt{\frac{3}
                                          {4}(\vec{l}^{\, 2}-\vec{k}^2)+M B}
                             -\sqrt{M B}}) } \label{fonzietilde}\\
      & & \phantom{\int \frac{d^3 l}
                 {(2\pi)^3}}
       \frac{1}
            {\vec{l}^2 -\vec{l}\cdot\vec{p}+\vec{p}^{\, 2}-\frac{3}
                                                           {4}\vec{k}^2+M B},
   \nonumber 
\end{eqnarray}
\noindent
where $B$ is the deuteron binding energy and $\tilde t$ is a correlator with
two dibaryon and two nucleons external legs represented by the blob in 
\ref{fig2}. One could have computed the deuteron-nucleon amplitude from 
the correlator involving any other operator with the deuteron quantum numbers
instead of the dibaryon used here. The results 
for the (on shell) deuteron-nucleon scattering are obviously independent 
of this choice
as long as the proper wave function renormalization is performed. 
The scattering matrix is given by
\begin{equation}
t(p)= \sqrt{Z}\  \tilde t(p,p)\  \sqrt{Z},
\end{equation}
where $Z$ is defined by
\begin{equation}
 Z =  i S_0(p)|_{\rm pole}/i S(p)|_{\rm pole}=
               -\frac{1}
                     {1-\frac{M^2g^2}
                             {4 \pi \sqrt{MB}}
                     }.
\end{equation}
\noindent
It is convenient then to define the off-shell scattering amplitude
\begin{equation}
t(p,k)=\frac{\vec{p}^{\, 2}-\vec{k}^2} 
            {-\frac{3(\vec{p}^{\, 2}-\vec{k}^2) }
                   {8 M^2 g^2}
              +\frac{1}
                    {4 \pi}({\sqrt{\frac{3}
                                          {4}(\vec{p}^{\, 2}-\vec{k}^2)+M B}
                             -\sqrt{M B}}) } \tilde t(p,k), 
\end{equation}
\noindent
satisfying
\begin{eqnarray}
\lefteqn{ \left[-\frac{3(\vec{p}^{\, 2}-\vec{k}^2) }
                {8 M^2 g^2}
           +\frac{1}
                 {4 \pi}({\sqrt{\frac{3}
                                          {4}(\vec{p}^{\, 2}-\vec{k}^2)+M B}
                             -\sqrt{M B}})\right]
 \frac{t(\vec{p}, \vec{k})}
      {\vec{p}^{\, 2}-\vec{k}^2-i \epsilon }}    \\ \label{aeq}
& & = \frac{-1}
           {(\vec{p}-\vec{k}/2)^2+M B}
       -   \int \frac{d^3 l}
                             {(2\pi)^3} 
       \frac{1}
            {\vec{l}^2 -\vec{l}\cdot\vec{p}+\vec{p}^{\, 2}-\frac{3}
                                                           {4}\vec{k}^2+M B}
         \frac{t(\vec{l}, \vec{k})}
              {\vec{l}^2-\vec{k}^{\, 2}-i\epsilon},  \label{fonzie}  \nonumber 
\end{eqnarray}
\noindent
that reduces to the scattering amplitude $t(p)$ on shell
\begin{equation}
t(p)=t(p,p).
\end{equation}
\noindent

Since we are interested only in s-wave scattering, we should project
this equation into its $L=0$ component. The result is
\begin{eqnarray}
\lefteqn{  \frac{3}{2}\left[ - \eta + \frac{1}{\sqrt{\frac{3}{4}
            (x^2-y^2)+1}+1}\right]
               a(x,y)=
-\frac{1}{xy}{\rm ln}\left(\frac{(x+y/2)^2+1}{ (x-y/2)^2+1}\right)}\nonumber\\
& & -\frac{2}{\pi x}\int_0^\infty \ dz\ z {\rm ln}
    \left(\frac{z^2+x^2+1 - \frac{3}{4} y^2 + xz}
               {z^2+x^2+1 - \frac{3}{4} y^2 - xz}\right) 
       \frac{a(z,y)}{z^2-y^2- i\epsilon},\label{xyz} 
\end{eqnarray}
\noindent
where we use the dimensionless quantities $x=p/\sqrt{MB}$, $y=k/\sqrt{MB}$, 
$z=l/\sqrt{MB}$, and 
$a(x,y)=\frac{\sqrt{ MB}}{4 \pi} 
t_{L=0}(p,k)$,
and $\eta=\sqrt{MB} r_0/2$.
For finite values of $k$ this equation is complex even below threshold 
($3 k^2/4 = B$) due to the $i\epsilon$ prescription. 
The numerical solution of the equation above is trivial in terms of 
computer power. The only subtle point is how to deal with the 
$i \epsilon$ prescription that appears in eq.(\ref{xyz}).
One way of dealing with that is  to use the real $K$-matrix defined by
\begin{equation}
K(x,y)=\frac{a(x,y)}{1+i y a(y,y)},
\end{equation}
\noindent
which satisfies the equation
\begin{equation}
K(x,y) = - h(x,y,y) - \frac{2}{\pi}\int_0^\infty\ dz\ z^2 h(x,y,z) 
        \frac{\cal P}{z^2-y^2}K(z,y),
\end{equation}
\noindent
with
\begin{eqnarray}
h(x,y,z)  &=& \frac{1}{x z \tilde f(x,y)} {\rm ln} \left(
\frac{z^2+x^2+1-\frac{3}{4}y^2+xz}{z^2+x^2+1-\frac{3}{4}y^2-xz}
\right),\nonumber \\
\tilde f(x,y) &=&\frac{3}{2}\left[ - \eta + \frac{1}{\sqrt{\frac{3}{4}
       (x^2-y^2)+1}+1}\right].
\end{eqnarray}
The phase shifts can be obtained directly form the on-shell $K$-matrix :
\begin{equation}
k {\rm cot}\delta=\frac{\sqrt{MB}}{K(\frac{k}{\sqrt{MB}},\frac{k}{\sqrt{MB}})}.
\end{equation}
\noindent
Defining $f(x,y)$ by the equation
\begin{equation}
f(x,y)=\frac{h(x,y,y)}{h(y,y,y)}  
 -\frac{2}{\pi}\int_0^\infty dz z^2
\left( h(x,y,z)-\frac{h(x,y,y)}{h(y,y,y)}h(y,y,z)\right) 
      \frac{f(z,y)}{z^2-y^2},
\label{eqf}
\end{equation}
\noindent
the on-shell $K$-matrix can be obtained by
\begin{eqnarray}
K(y,y)& &=-h(y,y,y)\label{eqK}\\
& &\left(1+\frac{2}{\pi}\int_0^\infty dz
                     \left(z^2 h(y,y,z)f(z,y)-
                           y^2 h(y,y,y)f(y,y)\right)\frac{1}{z^2-y^2}
                 \right)^{-1}.\nonumber  
\end{eqnarray}
\noindent

Rewriting Eq. (\ref{fonzie}) this way 
greatly simplifies its numerical solution,
for now the 
integrand is regular and the principal value can be
dropped from Eqs. (\ref{eqf}) and (\ref{eqK}).

\begin{figure}[htb]
\begin{center}
\psfig{figure=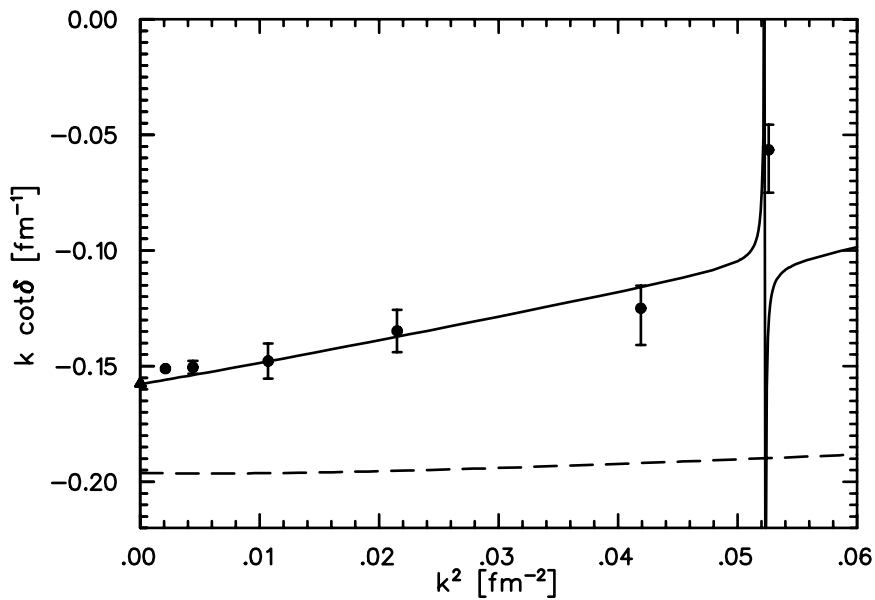,width=4.5in,angle=90}
\end{center}
\caption{$k\ {\rm cot}\delta$ in the $J=3/2$ channel to order 
$(r_0/a)^0$ (dashed line) and $(r_0/a)^2$ 
(solid line). Circles  are from the phase shift analysis in 
 \protect\cite{vanOers} and the triangle is from  \protect\cite{Dilg}.}
\label{fig3}
\end{figure}

We have solved \cite{threestooges} Eqs. (\ref{eqf}) and (\ref{eqK}) numerically and the result for 
the phase shifts for energies up to
the break-up point is shown in 
Fig. \ref{fig3}. The data points at finite energy were taken 
from the phase shift analysis in \cite{vanOers} and the much more precise 
(nearly) zero-energy point 
from \cite{Dilg}. Also plotted is the result of the leading order 
calculation
obtained by setting $\eta=0$,
in which case our equations reduce to the case studied in  \cite{skorny}.

We expect errors in our calculation to be of the order $(r_0/a)^3, (k r_0)^3$
compared to the leading order. These errors are smaller than the experimental
uncertainty in the finite energy case and of the same order as the 
experimental uncertainty in the case of the more precise measurement 
near $k=0$, 
where we find 
$^4a_{th}=6.33\pm 0.10$ fm \cite{genius}
compared to $^4a=6.35\pm 0.02$ fm \cite{Dilg}.
\vskip .4in

Our results seem to deviate from a simple effective range type expansion 
only around the pole at $\sim 0.05$ fm$^{-2}$ . 
(A pole in $k\ {\rm cot}\delta$ corresponds to a zero in the 
scattering matrix, which does not carry any special meaning.) 
This pole does not appear in potential model calculations
({\it e.g.}(\cite{jim})), and presumably will be smoothed out
by higher-order terms that we have not yet included. 
It is interesting that the only 
``experimental'' point in this region 
seems to indicate some structure there. More experimental information on 
this region would be highly desirable to confirm this prediction. More 
experimental points are available only beyond break up and are not shown
in Fig. \ref{fig3}. Notice that the break up region is still well within the 
range of validity of the pionless effective theory. It would be interesting
to check how well the predictions would compared to experiment in this region.

The calculation of higher-order corrections involves the knowledge of further
counterterms like the ones giving rise to p-wave interactions, etc. 
These
parameters can be determined either by fitting other experimental data
 or by matching with another effective theory
---involving explicit pions--- valid up 
to higher energies. 
If more precise experimental data ---particularly at 
zero-energy--- appear, we would be facing
a unique situation 
where precision 
calculations in strong-interaction physics can be 
carried out 
and tested \cite{ineedsomeonetohelpwannacollaborate?}. 
There is the prospect of better measurements at zero 
energy actually being carried out by using neutron interferometry 
techniques, so this is not
as far fetched a possibility as it may seem. 

\section{The $J=1/2$ channel}

The strong suppression of the three body force effects in the $J=3/2$ channel
is a reflection of the fact that, due to the Fermi principle, the 
three nucleons
cannot occupy the same position in space. In the $J=1/2$ channel, as well as 
in the case of three identical bosons, this restriction does not apply.
One might expect that the $J=1/2$ three body force term
does not have to include two derivatives like the $J=3/2$ three body force
does and that, as a consequence, its effect on  deuteron-nucleon scattering
is suppressed by only $(r_0/a)^4$ (as opposed to $(r_0/a)^6$ in the $J=3/2$
channel).  
There are however more profound differences between these channels.
Let us consider, for simplicity, a system of three identical bosons such that
there is a two boson bound state close to threshold. The leading order
contribution is given by the graphs in Fig.~\ref{fig2}, with $r_0=0$.
Power counting shows that all the graphs in this infinite series are 
ultraviolet finite. A closer look shows however that the sum of all graphs
contains some ultraviolet divergence. This has been known for a long time
and different aspects of this phenomenon lead to the so called Thomas 
\cite{thomas} 
and Efimov \cite{efimov} effects. Maybe the easier way to see 
this is to start from the
analogue of equation (\ref{fonzietilde}) corresponding to the bound state
of three bosons

 \begin{equation}
t(p,k)=\frac{2}{\pi} \int_0^\infty dq \ t(q,k) \frac{q}{p}
        \frac{1}{-\frac{1}{a}+\sqrt{\frac{3 q^2}{4}-ME}}
        {\rm ln}\left( \frac{p^2+q^2+p q - ME}{p^2+q^2-p q - ME}\right),
          \label{bosons}
\end{equation}
\noindent
where $E=-B+3 k^2/4M$ is the total energy.
Equation (\ref{bosons}) differs form (\ref{fonzietilde}) only in the absence of
the inhomogeneous term and in the coefficient in front of the kernel, that 
has a crucial opposite sign. We can map (\ref{bosons}) into a two dimensional
quantum mechanics problem through the transformation

\begin{equation}
\chi(r,\rho)=\int_0^\infty dp\  t(p,k) 
                     \frac{p \ {\rm sin}(\sqrt{3/4}p\rho)}
                          {-\frac{1}{a}+\sqrt{\frac{3 p^2}{4}-ME}}
               e^{-r \sqrt{\frac{3 p^2}{4}-ME}}.
\end{equation}
$\chi(r,\rho)$ satisfies
\begin{equation}
\left( \frac{\partial^2}{\partial r^2}+\frac{\partial^2}{\partial \rho^2}+
       ME\right)\chi(r,\rho)=0,
\end{equation}
\noindent
with the boundary conditions
\begin{eqnarray}
&&\chi(r,0)=0\\
&&\frac{\partial}{\partial r}\chi(r=0,\rho) - \frac{1}{a}\chi(r=0,\rho)
+\frac{8}{\sqrt{3}}\frac{1}{\rho}
           \chi(\frac{\sqrt{3}}{2}\rho,\frac{1}{2}\rho)=0.
\label{bc1}
\end{eqnarray}
\noindent
This same Schroedinger equation with these boundary equation would be obtained
by considering the three particle problem in the usual wave mechanics
formalism and using the 
Jacobi coordinates. The variable $r$ corresponds to the distance between 
two of the particles and $\rho$ to the distance between the third particle
and the center of mass of the other two. 
Using polar coordinates $R=\sqrt{r^2+\rho^2},{\rm tg}\alpha=r/\rho$,
we can see that this complicated boundary condition is simpler for
$R\ll a$. Equation (\ref{bc1}) becomes
\begin{equation}
\frac{\partial}{\partial\alpha}\chi(R,\alpha=0)+
\frac{8}{\sqrt{3}}\chi(R,\pi/3)=0
\label{bc2}
\end{equation}
\noindent
In the region $R\ll a$ we have then a separable problem with the solution
\begin{equation}    
\chi(R,\alpha)=\sum_i F_i(R) {\rm sin}(s_i(\pi/2-\alpha))
\end{equation}
\noindent
where the $s_i$ are the solutions of
\begin{equation}
s_i\ {\rm cos}(s_i \pi/2)=\frac{8}{\sqrt{3}}{\rm sin}s_i\pi/6 .\label{s0}
\end{equation}
\noindent
Equation (\ref{s0}) has one imaginary solution $is_0$, with $s_0\simeq 1.006$.
The equation for the radial part $F_0(R)$ contains an {\it attractive}
$1/R^2$ potential
\begin{equation}
\left(\frac{1}{R}\frac{d}{dR}R\frac{d}{dR}-\frac{s_0^2}{R^2}+ME\right)
    F_0(R)=0.\label{r2}
\end{equation}
\noindent
Equation (\ref{r2}) is valid for $1/\Lambda\ll R \ll a$, where $\Lambda$
is an ultraviolet cutoff. Taking 
$a\rightarrow 0$ with $\Lambda$ fixed we find that an infinite number 
of bound states will appear at threshold (Efimov effect). Keeping
$a$ fixed and taking $\Lambda\rightarrow\infty$ bound states  appear 
with arbitrarily low energy. That suggests that even at leading order
there is need for a counterterm at $r=\rho=0$ (three body force).
What is far from clear is whether only one counterterm would be enough 
to absorb the divergence or an infinite number of them 
(a whole form factor)
would be necessary even at leading order \cite{coolpaper}. 
Of course this last possibility
would destroy the predictive power of the effective theory program in 
nuclear physics, at least on schemes in which pion exchanges 
are perturbative.

Notice that the fact  that three body forces appear at leading order
in our approach 
does not contradict the usual statement that in phenomenological 
potential models
they are small. By changing the cutoff the effect of two body forces
can be  transfered to the three body forces (in other words, the beta 
function for the three body force depends on the two body force).

Another way of seeing that the ultraviolet problem described here cannot be
seen diagram by diagram is to multiply the kernel of (\ref{bosons}) by a 
parameter $\lambda$. Perturbation theory in $\lambda$ at any finite order
corresponds to 
the truncation of the series in Fig. \ref{fig2}. Repeating the arguments
of this section we now find that the value of $s_0$ depends on a non analytic
way on $\lambda$. $s_0$ is zero for $\lambda$ smaller than some critical value,
and becomes non zero for larger values of $\lambda$. Perturbation theory
around $\lambda=0$ will  see any no hint of a finite $s_0$ and no
attractive $1/R^2$ potential.


\section*{Acknowledgments}
The work presented in this talk was done in collaboration with 
H.~-W.~Hammer and U.~van Kolck. I thank D. Kaplan for extensive 
discussions on the subject.

This research was supported in part by the U.S. Department of Energy
grants DOE-ER-40561.

\end{document}